\documentclass[aps,pra,superscriptaddress,twocolumn,floatfix,showpacs]{revtex4-2}

\usepackage{graphicx}
\usepackage{amsmath}
\usepackage{amsfonts}

\usepackage{amsthm}
\usepackage{amssymb}
\usepackage{hyperref}
\usepackage[utf8]{inputenc}
\usepackage{mathtools}
\usepackage[english]{babel}
\usepackage{bbm}
\usepackage{float}
\hypersetup{colorlinks=true, linkcolor=blue, citecolor=blue, urlcolor=blue}

\usepackage{subfigure,upgreek}
\usepackage{xcolor}
\usepackage[normalem]{ulem}
\usepackage{braket}
\usepackage{microtype}
\usepackage{bbm}
\usepackage{color}
\usepackage{dsfont}
\usepackage{bbold}

\newcommand{\abs}[1]{\lvert #1 \rvert} 
\DeclareMathOperator{\Imag}{Im}
\DeclareMathOperator{\Real}{Re}

\begin{document}




\title{Harnessing quantum emitter rings for efficient energy transport and trapping}



\author{R. Holzinger}
\affiliation{Institute for Theoretical Physics, University of Innsbruck, Technikerstrasse 21a, A-6020 Innsbruck, Austria}
\author{J. S. Peter}
\affiliation{Department of Physics, Harvard University, Cambridge, Massachusetts 02138, USA}
\affiliation{Biophysics Program, Harvard University, Boston, Massachusetts 02115, USA}
\author{S. Ostermann}
\affiliation{Department of Physics, Harvard University, Cambridge, Massachusetts 02138, USA}
\author{H. Ritsch}
\affiliation{Institute for Theoretical Physics, University of Innsbruck, Technikerstrasse 21a, A-6020 Innsbruck, Austria}
\author{S. F. Yelin}
\affiliation{Department of Physics, Harvard University, Cambridge, Massachusetts 02138, USA}
\date{\today}

\begin{abstract}
Efficient transport and harvesting of excitation energy under low light conditions is an important process in nature and quantum technologies alike. Here we formulate a quantum optics perspective to excitation energy transport in configurations of two-level quantum emitters with a particular emphasis on efficiency and robustness against disorder. We study a periodic geometry of emitter rings with subwavelength spacing, where collective electronic states emerge due to near-field dipole-dipole interactions. The system gives rise to collective subradiant states that are particularly suited to excitation transport and are protected from energy disorder and radiative decoherence. Comparing ring geometries with other configurations shows that that the former are more efficient in absorbing, transporting, and trapping incident light. Because our findings are agnostic as to the specific choice of quantum emitters, they indicate general design principles for quantum technologies with superior photon transport properties and may elucidate potential mechanisms resulting in the highly efficient energy transport efficiencies in natural light-harvesting systems.
\end{abstract}

\maketitle

In quantum optics, ordered quantum emitter lattices with subwavelength spacing have emerged as a resourceful platform for near-term quantum technologies \cite{Bekenstein2020Metasurfaces,Rui2020Mirror,Manzoni_2018,Perczel2017Topological,Masson2020Atomic,Gutierrez2022Coherent,Jen2018Cooperative,Guimond2019Subradiant,Reitz2022Cooperative,Patti2021Array,Shahmoon2017Cooperative,Buckley2022Optimized,Patti_2021}.
Here long-range interactions between light-induced dipoles
lead to highly modified optical properties of the quantum emitter ensemble, including Dicke superradiance \cite{GROSS1982301} and the emergence of collective long-lived subradiant states \cite{Garcia2017Exponential,Needham2019Subradiance}. Applications range from single photon switch gates \cite{Cardoner2021Quantum} to enhanced single photon detection for biomedical applications \cite{Bettles2016Enhanced,Aslam2023Sensor} and topological edge state lasing \cite{parto2018edge,st2017lasing}.
Likewise, uncovering design principles underlying biological systems and applying this understanding to synthetic systems is crucial for near-term quantum technologies.
Ring geometries of quantum emitters promise to enhance single photon sensing, transport, storage, and light generation in engineered nanoscale systems \cite{Cardoner2019Subradiance,Needham2019Subradiance,Holzinger2020Laser,Cardoner2022Antenna}.
In photosynthetic energy transfer, as it occurs in nature, organisms utilize ring-shaped antennae that increase the photon scattering cross-section of a single reaction center: the site where photosynthesis takes place. This transfer process occurs at near unit efficiency, and understanding the mechanisms behind this remarkable feat is an outstanding scientific challenge \cite{Scholes2011Photosynthesis,Olaya2008Efficiency,Mattioni2021Design,Scholes2011Photosynthesis,Levi2015Quantum,Escalante2010LongRange,Gitt2011Quantum,Lee2007Quantum,Creatore2013Quantum,Strumpfer2012Quantum,Potocnik2018Superconducting,Fassioli2014Photo}.\\
 \begin{figure}[ht!]
\includegraphics[width=0.94\columnwidth]{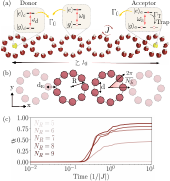}
\vspace{-0.9em}
\caption {\textbf{Lattices of nanoscopic quantum emitter rings.} \textbf{(a)} Each ring is composed of two-level quantum emitters with resonance frequency $\omega_0$ and separation $d<\lambda_0$, where $\lambda_0=\omega_0/c$ is the wavelength of light. The excited state $|e\rangle$ spontaneously decays with rate $\Gamma_0$ to the ground state $|g\rangle$ and the emitters are coupled via long-range dipole-dipole interactions with nearest-neighbor coupling strength $J$. Emitters acting as donor and acceptor are shown in yellow and the acceptor features an additional trapping state to which excitations irreversibly decay with rate $\Gamma_\mathrm{T}$. \textbf{(b)} More detailed sketch illustrating the inter-ring separation $d_R$. The ring radius $R$ and the emitter spacing $d$ are related via $d = 2R \sin(\pi/N_\mathrm{R})$, with $N_R$ emitters per ring. \textbf{(c)} Excitation transport efficiency according to Eq.~\ref{efficiency} for a chain of 10 rings and various $N_R$. Parameters: $d/\lambda_0 = 0.05$, $d_R/d=0.9$, $\Gamma_\mathrm{T}/\Gamma_0 = 2$ and $\Delta = 0$.}
\label{fig1}
\end{figure}
\indent Taking inspiration from biological systems, we examine the long-range excitation transport between a donor and an acceptor emitter through a lattice of quantum emitter rings [Fig. \ref{fig1}(a)]. As a main result, we show that efficient excitation transport at low trapping rates preferentially occurs for ring geometries, as compared to other lattices. This property has important consequences for devising artificial light harvesting and transport systems, and may be relevant for understanding the excellent excitation transport capabilities of biological systems \cite{mirkovic2017light}. We also highlight that for ring lattices, the trapping of light at an acceptor site under low-light conditions is enhanced by many orders of magnitude as compared to other geometries and to independent emitters. By choosing an optimal detuning for the donor and acceptor with respect to the lattice, radiative losses are strongly suppressed, and excitations are protected during the transport by subradiance \cite{Needham2019Subradiance,Garcia2017Exponential}. While the influence of the excitation trapping rate on the transport efficiency in other geometries has been explored in other works \cite{peter2023zeno, Soler2010Trapping,Reza2019Trapping,Timpmann2014Trapping}, as have certain design principles for bio-inspired artificial solar-harvesting devices \cite{Mattioni2021Design,Scholes2011Photosynthesis,Levi2015Quantum,Bredas2017Lessons,Kropf2019Transport,Akselrod2014Nanoscale,Semion2013Photonics,Celardo2014Robustness}, our findings specifically highlight the advantages of the rotationally symmetric ring geometry. This special feature of ring configurations
is particularly intriguing for its close connection to natural photosynthetic complexes found in biological systems. Our work therefore opens the possibility of exploiting quantum effects in bio-inspired configurations of quantum emitters for near-term optical technologies that enable quantum-enhanced light-matter coupling on the nanoscale. \\
\begin{figure*}[ht]
\includegraphics[width=1\textwidth]{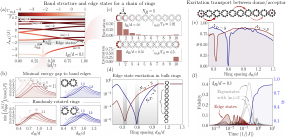}
\caption {\textbf{Band structure, edge states, and excitation transport for a chain of rings.} \textbf{(a)} Eigenmodes of the effective Hamiltonian $\hat{\mathcal{H}}_\mathrm{lattice}$ in Eq. (\ref{ham1}a) can be cast into $N_R=9$ energy bands with an angular momentum projection $|m|$ using Eq. (\ref{ansatz}) and translation along the ring chain axis is given by $\tilde{d}=2R+d_R$, the spacing between adjacent ring centers. (Decay rates of all eigenmodes are color coded). A band gap emerges with two edge states residing inside with an energy separation $\Delta_\mathrm{gap}^{(0,1)}$ to the nearest lower/upper band edge respectively. \textbf{(b)} The minimal energy gap $\Delta_\mathrm{gap}^{(0,1)}$ to the nearest lower/upper band edge normalized by the maximal nearest-neighbor coupling $J$ shows distinct maxima as a function of $d_R$. These maxima correspond to an optimal transport efficiency $\eta_t$ as shown in (e) for $N_R=8,9$. Furthermore, smaller emitter numbers per ring $N_R$ are effected more by randomly rotated rings except for even $N_R$. (The average was taken over 50 random realizations). \textbf{(c)} For $N_R=9$ edge states appear with maximum amplitude on the left/right end of the ring chain and a superradiant decay rate for decreasing inter-ring spacings $d_R/d$. Notably, topological edge states have been observed in zigzag chains of gold nano-rings  \cite{parto2018edge,st2017lasing} with lasing from the edge rings. \textbf{(d)} With decreasing inter-ring spacing $d_R$ edge states become more pronounced with distinct minima where the bulk amplitude vanishes leading to a suppression of excitation transport. \textbf{(e)} Excitation transport between a donor and acceptor site placed in the center of the edge rings. The transport efficiency $\eta_t$ is evaluated after a time $t \Gamma_0 = 150$ with the donor/acceptor detuning $\Delta$ optimized for maximum transport. Suppression appears when the edge states become too pronounced with vanishing amplitude in the bulk rings as shown in (d). \textbf{(f)} Time dynamics of the excitation transport process between a donor and acceptor site with $\Delta=0$. The eigenstate fidelity $|\langle \Psi(t)| \Psi_\mathrm{eig} \rangle |^2$ for $N_R=9$ demonstrates the importance of edge states at early times. Parameters: $d=0.05\lambda_0$, $\Gamma_\mathrm{T}/\Gamma_0 = 1$ in (e)-(f) and 10 rings in (b)-(f).}
\label{energy-band}
\end{figure*}
\indent As a paradigmatic quantum optical model to simulate excitation energy transport, we consider a one-dimensional lattice of $M$ rotationally symmetric rings, each composed of $N_R$ identical two-level emitters with a ground state $|g \rangle$ and an excited state $|e\rangle$. The two states are connected via the transition operator $\hat{\sigma}_n = |g_n \rangle \langle e_n|$ for the $n$-th emitter. Additional emitters acting as donor and acceptor sites are placed in the center of two rings at either end of the lattice, as illustrated in Fig. \ref{fig1}(a). The transport efficiency between these two sites is the core quantity of interest and is defined as
\begin{align} \label{efficiency}
\eta_t = \Gamma_\mathrm{T} \int_0^{t} dt' \langle \Psi(t')| \hat{\sigma}^\dagger_\mathrm{a} \hat{\sigma}_\mathrm{a}| \Psi(t')\rangle.
\end{align}
Here $t$ is the integration time over which excitation can accumulate in the trap state [see Fig. \ref{fig1}(a)]. $\eta_t$ can take values between 0 and 1, where 0 corresponds to no transport at all and 1 identifies maximal transport efficiency. $\hat{\sigma}^\dagger_a \hat{\sigma}_a = |e_a \rangle \langle e_a|$ corresponds to the projector onto the excited state of the acceptor.
The trap population accumulates over time and reaches a steady state value at large times $t$ when the total excited state population is either dissipated via radiative losses or accumulated in the trap. 
The transition frequencies and decay rates of the ring emitters are assumed to be equal and given by $\omega_0 = 2\pi c/\lambda_0$ and $\Gamma_0$, respectively, whereas the donor/acceptor transitions may be detuned by $\Delta=\omega_\mathrm{d,a}-\omega_0$ with respect to the ring emitter frequencies. We assume $\omega_\mathrm{d}=\omega_\mathrm{a}$ for the remainder of this work. The acceptor features an extra trapping channel through which excitations are extracted from the system at a rate $\Gamma_\mathrm{T}$. Furthermore, the quantum emitters are confined in the $x$-$y$ plane with intra-ring separation $d=2R\sin(\pi/N_R)$ and inter-ring separation $d_R$, where $R$ is the ring radius, as illustrated in Fig. \ref{fig1}(b). To reduce the number of free paramters, all dipole emitters are assumed to be circular polarized, namely $(1, i,0)^T/\sqrt{2}$. However, qualitatively similar results can be obtained for geometries consisting of linear polarized emitters.\\ 
\indent We model the system within the Born-Markov approximation \cite{Garcia2017Exponential}, and only consider the quantum emitter's internal degrees of freedom. Furthermore, we assume the weak excitation regime, where at most a single excitation is present in the system (see Methods), and therefore the system can be described (in the rotating frame with $\omega_0$) by the non-Hermitian Hamiltonian $\hat{\mathcal{H}}_\mathrm{eff} = \hat{\mathcal{H}}_\mathrm{ad}+\hat{\mathcal{H}}_\mathrm{lattice}+\hat{\mathcal{H}}_\mathrm{int}$. Here $\hat{\mathcal{H}}_\mathrm{ad} = (\Delta-\frac{i}{2}\Gamma_0) (\hat{\sigma}^\dagger_\mathrm{a}\hat{\sigma}_\mathrm{a}+\hat{\sigma}^\dagger_\mathrm{d} \hat{\sigma}_\mathrm{d}) - \frac{i}{2}\Gamma_\mathrm{T} \hat{\sigma}^\dagger_\mathrm{a}\hat{\sigma}_\mathrm{a}$ is the bare Hamiltonian of the donor and acceptor, $\hat{\mathcal{H}}_\mathrm{lattice}$ describes the emitters in the ring lattice, and $\hat{\mathcal{H}}_\mathrm{int}$ describes the interaction between the ring emitters and  the donor/acceptor,
\begin{subequations}
\begin{align}
\hat{\mathcal{H}}_\mathrm{lattice} &=\sum_{n,m} \Big(J_{nm}-i\frac{\Gamma_{nm}}{2}\Big) \hat{\sigma}^\dagger_n \hat{\sigma}_{m}, \\
\hat{\mathcal{H}}_\mathrm{int} &=\sum_{n;k=\mathrm{a},\mathrm{d}} \Big(J_{nk}-i\frac{\Gamma_{nk}}{2}\Big) (\hat{\sigma}^\dagger_n \hat{\sigma}_{k}+\hat{\sigma}^\dagger_k \hat{\sigma}_{n}). 
\end{align}
\label{ham1}
\end{subequations}
All emitters interact via vacuum-mediated dipole-dipole interactions in free space. The pairwise coherent and dissipative interactions are given by $J_{nm} = -3\pi \Gamma_0 /k_0 \Real (G_{nm})$ and $\Gamma_{nm}=6 \pi \Gamma_0 /k_0 \Imag (G_{nm})$ respectively, with $G_{nm}$ being the free space Green's function (see Methods). The Green's function depends only on the separations between the emitters and their dipole orientation.
The time evolution of the system is described by the effective Hamiltonian in Eq.~\eqref{ham1} via the Schrödinger equation $i \partial_t | \Psi(t) \rangle = \hat{\mathcal{H}}_\mathrm{eff}| \Psi(t) \rangle $. Since the Hamiltonian is non-Hermitian, the amplitude of the wavefunction for the quantum emitters decreases with time, which is a direct manifestation of the dissipative nature of the system. 

As discussed in previous works \cite{Cardoner2019Subradiance, Garcia2017Exponential,Needham2019Subradiance}, a subwavelength-spaced ring of quantum emitters exhibits guided eigenmodes that
are extremely subradiant, exhibiting an exponentially increasing lifetime, $\tau \Gamma_0 \sim \mathrm{exp}(N_R)$, of a single excitation \cite{Cardoner2019Subradiance}. Aside from the bright symmetric superposition state, the fields of the remaining eigenmodes vanish at the center of the ring due to symmetry. Thus, they are decoupled from any emitter at the center. Here we demonstrate that a donor/acceptor at the center of the ring that is dipole-dipole coupled to the symmetric ring mode can form a subradiant state with a majority of the excitation concentrated in the donor/acceptor. We start by analyzing a single ring of $N_R$ emitters with a single donor in the center. For a single ring, where the dipole orientations preserve the discrete rotational invariance, the collective eigenmodes of the effective Hamiltonian are spin waves of the form
$|\Psi_m \rangle = \hat{S}_m^\dagger |G\rangle$, where
\begin{align} \label{spinwaves}
\hat{S}_m = \frac{1}{\sqrt{N_R}}\sum_{j=1}^{N_R} e^{i m \varphi_j} \hat{\sigma}_j
\end{align}
and $|G\rangle$ denotes all emitters in the ground state. Here $\varphi_j = 2 \pi j/N_R$ is the angle between neighboring emitters along the ring and $m=0,\pm 1,\cdot \cdot \cdot,\lceil \pm (N_R-1)/2 \rceil$ is the angular momentum of the collective mode. The associated energy shifts and decay rates of these spins waves are given by $\tilde{J}_m= \sum_j e^{i m \varphi_j} J_{1j}$ and $\tilde{\Gamma}_m = \sum_j e^{i m \varphi_j} \Gamma_{1j}$, respectively. In such a configuration, all ring emitters couple equally to the central donor, which restricts the spectrum to the $m=0$ mode. This system features two eigenstates $|\Psi_\pm \rangle$, that are symmetric/anti-symmetric superpositions of the symmetric ring mode and the central donor. The anti-symmetric state can be extremely subradiant depending on the detuning $\Delta$ of the donor with respect to the ring emitters, resulting in a vanishingly small net dipole strength \cite{Cardoner2022Antenna}. This leads to an optimal detuning $\Delta_\mathrm{sub} \approx J_\mathrm{d}(\tilde{\Gamma}_0 -\Gamma_0)-\tilde{J}_0$ that maximizes the subradiance of the donor with an effective decay rate $\Gamma_\mathrm{eff}/\Gamma_0 \lesssim 10^{-3}$ (see Methods). Here, $J_\mathrm{d}$ is the coherent coupling between the donor and a ring emitter.
\newline
\indent Likewise, a chain of quantum emitter rings features a rich collective eigenmode structure. In particular, the subradiance of the eigenmodes protects the excitations from radiative decoherence and leads to efficient excitation transport \cite{Needham2019Subradiance,Ballantine2020Subradiance}.
As shown above, eigenmodes of a rotationally symmetric ring carry angular momentum $m$. Similarly, the eigenmodes of a linear chain of quantum emitters carry linear momentum $k$ \cite{Garcia2017Exponential,Masson2020Atomic}. This leads to an ansatz wavefunction for the eigenmodes of a ring chain, $|\Psi_{m,k} \rangle$, with an angular and linear quasi-momentum pair $(m,k)$, associated eigenenergies $\omega_0+J_\mathrm{m,k}$ and decay rates $\Gamma_\mathrm{m,k}$ \cite{Albrecht_2019} (details provided in the Methods). The translational distance between adjacent ring centers along the chain is given by $\tilde{d}=2R+d_R$. Fig. \ref{energy-band}(a) shows the energy bands for $N_R=9$ and $d_R/d = 0.9$. The band structure exhibits a nontrivial topology with a non-zero Zak phase $\varphi = i \int_{BZ} d k \, \langle \Psi_{m=0,k} | \partial_k| \Psi_{m=0,k} \rangle$ \cite{bernevig2013topological} as well as gapped edge states between the energy bands of the $m=0$ and $|m|=1$ eigenmodes. The edge states emerge for decreasing inter-ring spacings $d_R$, illustrated in Fig. \ref{energy-band}(c), and become more pronounced until a critical spacing of $d_R/d\approx 0.58$ and $d_R/d \approx 0.34$ for $N_R=8,9$ emitters per ring, respectively. The edge states are energetically degenerate and detuned by $\Delta_\mathrm{gap}^{(0,1)}$ from the lower/upper band edge respectively. Fig. \ref{energy-band}(b) shows the minimum distance of the edge states to the nearest band edge as a function of the inter-ring spacing $d_R$. Topologically protected edge states are crucial for resilient excitation transport in disordered systems \cite{Perczel2017Topological} and $\mathrm{min}(\Delta_\mathrm{gap}^{(0,1)})$ can serve as a figure of merit in this regard. Specifically for lattices of rings, the band gap remains finite in the presence of rotational disorder and exhibits a distinct maximum, e.g. $d_R/d \approx 0.9$ for $N_R=9$, as shown in Fig. \ref{energy-band}(b). Edge states also become superradiant at the critical distance where excitation transport is surpressed, as is shown in Fig. \ref{energy-band}(c). This points to the possibility of edge mode lasing, already observed in gold nano-rings arranged in zigzag chains \cite{parto2018edge,st2017lasing,ota2020active}.\\
\indent Figs. \ref{energy-band} (e) and (f) demonstrate the fundamental influence of the edge states on the transport dynamics between a donor and acceptor site for 10 rings with $N_R=9$. At the critical spacings, transport is either completely or strongly suppressed because the edge states possess no amplitudes in the bulk rings. Conversely at other spacings $d_R$, edge states are crucial during the early times of the transport process, as demonstrated by the eigenstate fidelities $\mathcal{F}(t) = |\langle \Psi(t)| \Psi_\mathrm{eig} \rangle |^2$ for $N_R=9$. Qualitatively similar results hold for other $N_R$. Indeed, edge states have been thoroughly studied in dimerized chains ($N_R=2$), which reproduces a long-range generalization of the well-known Su–Schrieffer–Heeger (SSH) model \cite{st2017lasing}.
A more complete discussion of the emergence of edge and corner states \cite{ota2020active} in two-dimensional ring lattices is briefly discussed in Methods and warrants further study.
 \begin{figure}[ht]
\includegraphics[width=1\columnwidth]{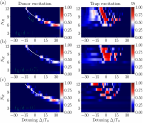}
\caption {\textbf{(a)-(c) Excitation transport in quantum emitter rings between a donor and acceptor site.} Scan over the number of emitters per ring $N_R$ and donor/acceptor detuning $\Delta$ for 10 rings with decreasing inter-ring spacings $d_R$ after a time $t \Gamma_0 = 150$. Efficient transport emerges only with $N_R \ge 6$, irrespective of $d_R$. The excited state population in the donor can get trapped in a subradiant state involving the ring surrounding the donor and follows the detuning $\Delta_\mathrm{sub}$ (white dashed line) derived in the main text. For 9-fold symmetric rings the donor/acceptor detuning, that optimizes transport, is given by $\Delta \approx 0$ for all inter-ring spacings. Additional parameters are: $\Gamma_\mathrm{T} = 2 \Gamma_0$, $d=0.05\lambda_0$.}
\label{fig4}
\end{figure}

\begin{figure*}[ht]
\includegraphics[width=1\textwidth]{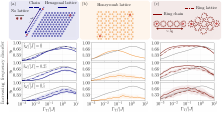}
\caption {\textbf{Excitation transport in ring geometries exhibits superior robustness against disorder.} Comparison of transport efficiency in \textbf{(a)} chain and hexagonal lattices and in the absence of a lattice (grey line), \textbf{(b)} honeycomb, and \textbf{(c)} ring lattices as a function of the trapping rate $\Gamma_\mathrm{T}$ and frequency disorder. Lattice emitter frequencies are randomly fluctuating by $\delta \omega$ around the resonance frequency $\omega_0$. The donor-acceptor distance is approximately the wavelength of light $\lambda_0$. Also shown in the dashed-dotted lines is the case of a donor-acceptor pair separated by $d$ in the absence of any lattice. Although frequency disorder decreases the long-range transport capacity, it prevails remarkably well even at large frequency fluctuations. In particular, the ring lattices exhibit high transport efficiencies (close to 90$\%$) over a wide range of trapping rates as compared to the other geometries. At trapping rates much below the magnitude of the coherent transfer rate $J$, ring-based lattices are superior to any other lattice in our study.
Additional parameters: $d_R/d= 0.9 $, $d/\lambda_0 = 0.06$, $J/\Gamma_0 \approx -8.4$, $t\Gamma_0=150$. An average over 25 random realizations with standard deviation $\delta \omega$ was performed in all plots. Donor/acceptor detunings in (a) $\Delta = 0$, (b) $\Delta = 4.5 \Gamma_0$, and (c) $\Delta=-\Gamma_0$.}
\label{fig6}
\end{figure*}
We now focus on the excitation transport dynamics and discuss the time evolution of a single initially excited donor. We find that excitation transport is optimized at particular donor/acceptor detunings, and that efficient transport occurs only for ring emitter numbers $N_R \ge 6$. In particular, rings with 8-, 9- and 10-fold symmetry seem to be most optimal. This is particularly intriguing because \cite{scheuring2009atomic,cleary2013optimal} 8-, 9- and 10-fold rings, the most abundant type occuring in natural light harvesting antennae \cite{Hu1998Architecture}, show the highest resilience when rings are randomaly rotated with respect to each other.
In Fig. \ref{fig4} the donor excited state populations and the trap populations $\eta_t$ are shown after a time $t \Gamma_0 = 150$ for a chain of 10 rings with various inter-ring spacings $d_R$. 
The detuning $\Delta$ where the donor excited state population is maximzed (i.e., most subradiant), follows the optimal detuning $\Delta_\mathrm{sub}$ for the single ring case. Here, the donor excitation largely remains trapped in the subradiant state discussed above, even for small inter-ring spacings.

\begin{figure*}[ht]
\includegraphics[width=1\textwidth]{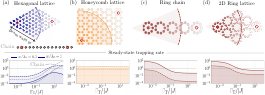}
\caption {\textbf{Ring lattices exhibit the most efficient energy trapping under weak light illumination and small trapping rates.} A comparison of the steady state effective trapping rate $\Gamma_\mathrm{T} \langle \hat{\sigma}_a^\dagger \hat{\sigma}_a \rangle_\mathrm{st} / (4\Omega_0^2)$ between different lattice geometries under continuous coherent driving with rate $\Omega_0 / \Gamma_0 = 10^{-3}$. The drive is modeled by a Gaussian beam with waist $w$ centered at the acceptor emitter and is on-resonance with the lattice emitter frequencies $\omega_0$ (see Eq. \ref{laser}). The effective trapping rate is normalized by the single emitter trapping rate driven on resonance (see main text). We study a hexagonal and chain lattice in \textbf{(a)} with $20$ and $8 \times 9$ emitters respectively, a honeycomb lattice in \textbf{(b)} with 130 lattice emitters, a chain of 5 rings in \textbf{(c)}, and a $3 \times 3$ hexagonal ring lattice in \textbf{(d)}. The donor/acceptor distance is approximately the wavelength of light $\lambda_0$ in all lattices. Strikingly, only the ring-based lattices are orders of magnitude more efficient at trapping incoming light for trapping rates below the nearest-neighbor coupling rate, namely when $\Gamma_\mathrm{T} \ll |J|$. Conversely, the honeycomb and hexagonal lattices are orders of magnitude less efficient in the same regime with the honeycomb lattice not deviating significantly from the single acceptor case. The linear chain (grey dashed) and the free space transfer (solid grey) shown in \textbf{(a)} is more efficient at trapping rates $\Gamma_\mathrm{T} > |J|$. Additional parameters: $d/\lambda_0 = 0.06$, $J/\Gamma_0 \approx -8.4$ and donor/acceptor detunings in \textbf{(a)} $\Delta/\Gamma_0 = -18$ (hexagonal), $\Delta = 0$ (chain), \textbf{(b)} $\Delta/\Gamma_0 = -20$, \textbf{(c)} $\Delta/\Gamma_0 = -3.85$, \textbf{(d)} $\Delta/\Gamma_0 =-4.63$ with $d_R/d = 0.9, \ N_R=9$, in \textbf{(c)} and \textbf{(d)}.}
\label{fig7}
\end{figure*}
A crucial element of excitation energy transport is robustness against energy disorder. We provide a comparison of ring lattices with other lattice geometries, including the influence of static frequency disorder in the lattice emitters. This is achieved by taking emitter frequencies $\omega_m$ from a Gaussian distribution around the unperturbed emitter frequency $\omega_0$ with a standard deviation $\delta \omega$ and adding the term $\sum_m (\omega_m-\omega_0) \hat{\sigma}^\dagger_m \hat{\sigma}_m$ to the Hamiltonian in Eq.~\eqref{ham1}. The donor/acceptor detuning $\Delta$ remains unchanged and is chosen such that unperturbed excitation transport is maximized. Fig. \ref{fig6} shows various geometries, many of which also have been studied previously for atoms trapped in optical lattices \cite{Masson2020Atomic, Patti2021Array,Buckley2022Optimized,Ballantine2020Subradiance, Perczel2017Topological, Cardoner2019Subradiance}. The nearest-neighbor distance is kept at $d=0.06 \lambda_0$ and the donor-acceptor distance at $\sim \lambda_0$ to establish a uniform comparison between the different geometries. Figs. \ref{fig6}(a) and \ref{fig6}(b) show a hexagonal lattice with $\Delta=0$ and a honeycomb lattice with $\Delta = 4.5 \ \Gamma_0$. In Fig. \ref{fig6} (c) a 1D ring chain and 2D hexagonal ring lattice with $N_R=9$ are shown with $\Delta= \Gamma_0$. The fluctuation in the emitter frequencies $\delta \omega$ is set to $|J|/4$ and $|J|/2$, where $J \approx -8.4 \Gamma_0$. Altogether, the different geometries show a similar reduction in the maximal transport efficiencies under disorder but behave quite differently in the range of trapping frequencies $\Gamma_\mathrm{T}$ where maximal transport occurs. Whereas the hexagonal lattice exhibits peak transport at a trapping rate far above the optical decay rate, namely at $\Gamma_\mathrm{T}/|J| \sim 2 $, the ring lattices demonstrate efficient transport over a large range $\Gamma_\mathrm{T}^\mathrm{opt}/|J| \sim 0.01-1$, even in the moderately disordered case. Qualitatively similar conclusions also apply to ring lattices with $N_R \neq 9$. Just as importantly, the ring lattices in Fig. \ref{fig6}(c) show significant transport enhancement for $\Gamma_\mathrm{T} \ll |J|$ compared to the independent case where no lattice is present. In summary, the ring lattices show significantly better transport capability and robustness against disorder.\\
\indent So far we have assumed, that a single donor is initially excited, and we have quantified the transport behavior by calculating the fraction of the excitation that accumulates in a trap state via Eq.~\eqref{efficiency} after a waiting time $t$. However, in many realistic scenarios, a perfectly excited donor is rather unlikely, and emitters close to the donor will be excited too. This motivates the study of the trapping rate at which the excitation ends up in the trap state under continuous coherent illumination in the form of a Gaussian laser beam with finite beam waist $w$. The continuous coherent drive is modeled
by
\begin{align} \label{laser}
    \hat{\mathcal{H}}_\mathrm{laser} = \Omega_0 \sum_i \exp{\Big(-\frac{|\vec{r}_d-\vec{r}_i|^2}{2w^2}\Big)}(\hat{\sigma}^\dagger_i+\hat{\sigma}_i),
\end{align}
where $\Omega_0$ is the laser Rabi frequency, $\vec{r}_d$ is the position of the donor with the sum including all emitters.
The driving rate of the laser is kept small ($\Omega_0 \ll \Gamma_0$) to ensure that the system stays in the single-excitation regime and the model remains valid. 
As a figure of merit for energy transport efficiency, we define $\Gamma_\mathrm{T} \langle \hat{\sigma}_a^\dagger \hat{\sigma}_a \rangle_\mathrm{st} / (4\Omega_0^2)$, as the steady-state trapping rate at the acceptor emitter.
The effective trapping rate is normalized by the trapping rate of a single acceptor, given by $\sigma_0 \Gamma_0 \Gamma_\mathrm{T}/(\Gamma_0+\Gamma_\mathrm{T})^2$, where $\sigma_0=6\pi/k_0^2$ is the single emitter scattering cross-section \cite{Cardoner2022Antenna}.
In Fig. \ref{fig7} a hexagonal lattice (a), a honeycomb lattice (b), and a hexagonal ring lattice with $N_R=9$ (c) are compared for two laser beam waists under continuous driving. By choosing a beam waist of $w/\lambda_0=0.3 $, most of the incoming light is focused around the donor emitter while the acceptor emitter remains mostly undriven. For $w/ \lambda_0=3$ the whole lattice is uniformly driven---a scenario more applicable to deeply subwavelength lattices under illumination from a non-directional light source. Natural light-harvesting antennae in purple bacteria offer an example \cite{Scholes2011Photosynthesis}.
In all cases, the donor/acceptor detuning $\Delta$ is chosen optimally such that the trapping rate is maximized. We find that the ring lattice is many orders of magnitude more efficient in trapping incident light as compared to both the triangular and honeycomb lattices as well as the single emitter at trapping rates much below the nearest-neighbor coherent transfer rate $J$. In particular for $\Gamma_\mathrm{T}/|J| \lesssim 0.01$ the ring lattice exhibits an almost $ 100 \times$ higher trapping efficieny compared to an independent emitter and the other lattices.

In conclusion, we have demonstrated intriguing optical properties of quantum emitter ring lattices, including the emergence of topological edge states. Furthermore, we have shown based on general symmetry principles that ring lattices form a superior platform for transporting and trapping excitations. We have also elucidated the guiding principles that govern optimal donor/acceptor detunings, trapping rates, and geometric arrangements with robustness against static energy disorder. Under more realistic conditions of weak coherent light illumination, we have shown that ring lattices are orders of magnitude more efficient at trapping the absorbed light when the trapping rate is much smaller than the nearest-neighbor coherent coupling rate. This result is thought-provoking since natural light-harvesting systems also operate with trapping mechanisms that are orders of magnitude slower than the coherent transfer time between neighboring chromophores \cite{chan2018single,Rienk1994energy}. Measurements performed on natural light-harvesting complexes show that the coherent energy transfer time between neighboring chromophores during photosynthesis is of the order of $\sim 0.1-10$ ps whereas the trapping time in the reaction center is typically $\sim 0.1-10$ ns \cite{Rienk1994energy,mirkovic2017light,chan2018single}. For a pre-existing trapping structure, this could provide an explanation why nature utilizes ring geometries as a moderating mechanism to trap absorbed sun light in reaction centers.
Other studies have focused on molecular emitters in ambient conditions with vibrational degrees of freedom and multiple decoherence channels \cite{Mohseni2008Transfer,Plenio_2008}. The impact of these additional effects on the results presented here are an exciting avenue for future research \cite{Cao2020Biology,Sohang2022Efficient,Palecek2017Coherence}.
Nevertheless, our results suggest that there exist general and platform-agnostic design principles that govern the efficient transport of excitation energy at the nano scale. These geometrical considerations may have played a role in evolutionary design and warrant further study.

\textbf{Acknowledgments} -- R.~H. acknowledges funding from the Austrian Science Fund (FWF) doctoral college DK-ALM W1259-N27.  S.O. is supported by the Harvard Quantum Initiaive (HQI). S.F.Y. thanks the AFOSR
and the NSF (through the CUA PFC and QSense QLCI).

\onecolumngrid


\section*{Methods}
\section*{Free space Green's function} 

The free-space Green’s function $\textbf{G}(\textbf{r},\textbf{r'}, \omega_0)$ is the solution to the electromagnetic wave equation in vacuum, and obeys
\begin{align}
\pmb{\nabla} \times \pmb{\nabla} \times \textbf{G}(\textbf{r},\textbf{r'}, \omega_0) - \frac{\omega_0^2}{c^2}  \textbf{G}(\textbf{r},\textbf{r'}, \omega_0) = \delta (\textbf{r}-\textbf{r'})  \mathbb{1}_3,
\end{align}
where $\mathbb{1}_3$ is the identity matrix in Cartesian coordintes. The Green’s function describes the field at point
$\textbf{r}$ due to a normalized, oscillating dipole at $\textbf{r'}$. The emitters are assumed to be fixed in space, allowing a treatment of the emitters as quantum
point dipoles instead of dynamical variables. Writing $\textbf{G}(\textbf{r}_n,\textbf{r}_m, \omega_0) = \textbf{G}(\textbf{r}_{nm},\omega_0)$, the Green's function is given by 
\begin{align}
\textbf{G}(\textbf{r}_{nm}, \omega_0) = \frac{e^{i k_0 r_{nm}}}{4\pi k_0^2 r^3_{nm}} \Big[(k_0^2 r_{nm}^2+i k_0 r_{nm}-1)\mathbb{1}_3 -(k_0^2 r^2_{nm}+3i k_0 r_{nm}-3) \frac{\pmb{r}_{nm}\otimes \pmb{r}_{nm}}{r_{nm}^2} \Big],
\end{align}
where $r_{nm} = \abs{\mathbf{r}_{nm}}$ and $k_0=2\pi/\lambda_0 = \omega_0/c$ is the wave number corresponding to the atomic transition energy, and $c$ is the speed of light in free-space.
The rates for coherent and dissipative interactions between emitter $n$ and $m$ are given respectively by
\begin{align} \label{greenstensor1}
J_{nm} &= - \frac{3\pi \Gamma_0}{k_0}  \Real (G_{nm}) \\
\Gamma_{nm} &= \frac{6\pi \Gamma_0}{k_0} \Imag (G_{nm}), \label{greenstensor2}
\end{align}

where we have defined $G_{nm} = \pmb{p}^{*}_n \cdot  \mathbf{G}(\mathbf{r}_{nm}, \omega_0) \cdot \pmb{p}_m$ and $\pmb{p}_n$ is the transition dipole matrix element vector for atom $n$, which we set to $\pmb{p}_n=(1,i,0)^T/\sqrt{2}$ (circular polarization) througout the main text. For a single emitter, evaluating Eq. \ref{greenstensor2} reproduces the vacuum emission rate $\Gamma_{nn}=\Gamma_0$ , where $\Gamma_0 = \omega_0^3 |\pmb{\mu}|^2/3\pi \hbar c^3$, and $|\pmb{\mu}|$ is the dipole moment strength. The single-emitter energy shift $J_{nn}$ in Eq. \ref{greenstensor1} formally yields a divergence
and is set to zero, as it would simply renormalize the resonance frequency $\omega_0$.

\section*{Wavefunction ansatz for a ring chain}

Here we show the wavefunction ansatz that is used to create the energy band structure.
As shown in the main text, one can assign an angular momentum number $m$ to each eigenmode of a rotationally symmetric ring. Similarly, the eigenmodes of a linear 1D chain of quantum emitters carry linear momentum $k$ \cite{Garcia2017Exponential}. This suggests, that the eigenmodes of a chain of rings can be understood in terms of wavepackets with a combination of angular and linear momentum $(m,k)$, which thus serves as a label of the eigenmodes \cite{Albrecht_2019}.
The ansatz is given by $|\Psi_{m,k} \rangle = \hat{S}_{m,k}^\dagger |g\rangle$ with
\begin{align} \label{ansatz}
\hat{S}_{m,k} = \frac{1}{\sqrt{N}} \sum_{j=1}^{N_R} \sum_{j'=0}^{M-1} e^{i m \varphi_j} e^{ i k \tilde{d}j'} \hat{\sigma}_{j+j' N_R},
\end{align}
where $N=M \times N_R$, $M$ is the number of rings, $\varphi_j = 2\pi (j-1)/N_R$, and $\tilde{d} = 2R+d_R$ is the distance between neighboring ring centers. 
To order the eigenmodes into energy bands with linear momentum $k$ and angular momentum $m$, we calculate the overlap fidelities of the exact numerical eigenstates $|\Psi \rangle$ of $\hat{\mathcal{H}}_\mathrm{lattice}$ with the ansatz in Eq. \ref{ansatz}, namely $|\langle \Psi_{m,k}| \Psi \rangle |^2$. The fidelities exhibit peaks centered around a particular linear momentum $k$ and angular momentum projection $|m|$ the numerical eigenmodes can then be ordered into energy bands. The ansatz only allows us to distinguish energy bands with angular momentum projection $|m|$ because the ansatz cannot discriminate between $\pm m$ except for $m=0$, which is unique. Alternatively, one can obtain the energy bands for an infinite chain of rings by block diagonalizing $\hat{ \mathcal{H}}_\mathrm{lattice}$, taking the individual rings as unit cells and invoking Bloch's theorem \cite{Perczel2017Topological,peter2023chiralityinduced}. However, here the bands cannot be split into their angular momentum projection, since the individual rings are numerically diagonalized. 

\section*{Ring with a central emitter}

Subwavelength rotationally symmetric rings of emitters feature eigenmodes that are extremely subradiant, but whose fields vanish at the center of the ring due to rotational symmetry. They are thus decoupled from any emitter at the center. The center only couples to the symmetric ring mode and the effective Hamiltonian can be rewritten as
\begin{align} \label{appB_ham}
\hat{\mathcal{H}}_\mathrm{eff} = \Big(\Delta-\frac{i\Gamma_0}{2}\Big)\hat{\sigma}_\mathrm{d}^\dagger \hat{\sigma}_\mathrm{d} + \Big(\tilde{J}_0-\frac{i\tilde{\Gamma}_0}{2}\Big) \hat{S}_0^\dagger \hat{S}_0 + \sqrt{N_R}\Big(J_d-\frac{i\Gamma_d}{2}\Big)(\hat{S}_0^\dagger \hat{\sigma}_\mathrm{d} + \hat{S}_0 \hat{\sigma}_\mathrm{d}^\dagger),
\end{align}
where $\tilde{J}_0$ and $\tilde{\Gamma}_0$ are the collective couplings of the symmetric ring mode, ${J}_d$ and ${\Gamma}_d$ are the couplings between the central donor and a ring emitter and $\Gamma_0$ is the spontaneous decay rate of a single emitter.
The field created by the symmetric ring mode at the donor position is the same as that of a single dipole with dipole moment strength enhanced by a factor of $\sqrt{N_R}$.
For the single excitation manifold, one finds the
eigenmodes by diagonalizing the $2 \times 2$ matrix resulting
from projecting Eq. \ref{appB_ham} into the subspace spanned by the basis $\{|\Psi_0 \rangle, |d\rangle \}$, with $|\Psi_0\rangle = S_0^\dagger |G\rangle$ and $|d\rangle = \hat{\sigma}_d^\dagger |G\rangle$. The corresponding eigenvalues are given by

\begin{align}
\lambda_{\pm}(\Delta) = \frac{1}{2}\Bigg(\tilde{J}_0-\Delta-\frac{i}{2}{(\tilde{\Gamma}_0+\Gamma_0)} \mp \sqrt{\Big[\tilde{J}_0+\Delta-\frac{i}{2}{(\tilde{\Gamma}_0-\Gamma_0)} \Big]^2 + 4 \Gamma_0 N_R \Big[J_d-\frac{i}{2}{\Gamma_d} \Big]^2}\Bigg).
\end{align}

For $d \le \lambda_0/3$ the eigenvalue $\lambda_- (\Delta)$ corresponds to the asymmetric superposition eigenmode between donor and symmetric ring mode with a majority of the excitation concentrated in the donor. Varying the donor detuning $\Delta$ varies the decay rate of the eigenmode, given by $-2 \ \mathrm{Im} \lambda_- (\Delta)$, and leads to a detuning $\Delta_\mathrm{sub}$ that minimizes the decay rate. For $d \ll \lambda_0$, the optimal detuning can be approximated by
\begin{align}
\Delta_\mathrm{sub} \approx J_d(\tilde{\Gamma}_0-\Gamma_0)-\tilde{J}_0.
\end{align}
For $N_R=9$, $\tilde{J}_0 \approx (N_R-1) J_d$ which yields $\Delta_\mathrm{sub} = 0$.
For a small total dipole moment, the central donor needs to inherit a larger fraction of the excitation, since the symmetric ring mode features a collective diple moment that is enhanced by a factor of $\sqrt{N_R}$. In the ideal Dicke case with $J_{nm} = 0$ and $\Gamma_{nm}=\Gamma_0$ for all $n,m$, the excitation fraction in the donor is given by $\langle \Psi_- | \hat{\sigma}_\mathrm{d}^\dagger \hat{\sigma}_\mathrm{d}| \Psi_-\rangle = {N_R}/({N_R+1})$, while the excitation fraction in each of the ring emitters is given by $ 1/(N_R(N_R+1))$. 
The anti-symmetric state is given by $| \Psi_- \rangle = (\sqrt{N_R} \hat{\sigma}^\dagger_\mathrm{d} - \hat{S}^\dagger_{0}) |g\rangle/\sqrt{N_R+1}$, but due to finite emitter separations and contributions from the dipole-dipole energy shifts, the actual anti-symmetric state deviates from the ideal case \cite{Cardoner2022Antenna}. This ideal scenario can be realized experimentally, for instance with artificial atoms interacting via a single-mode 1D waveguide \cite{Holzinger2022Localized}, but without transport capabilities due to the absence of coherent interactions.

\section*{Role of the trapping rate} 

Here we briefly discuss the influence of the trapping rate $\Gamma_T$ on excitation transport (for a more complete analysis, see ref. \cite{peter2023zeno}). We show that a large enough trapping rate leads to suppression of excitation transport, exhibiting the well-known quantum Zeno effect \cite{facchi2009quantum}. Conversely, a trapping rate that is too small reduces excitation transport as well, leading to an optimal rate $\Gamma_\mathrm{T}$. Furthermore, we show that the optimal trapping rate is closely related to the group velocity $v_g(k) = |\partial_k J_{m,k}|$ of particular collective modes in a specific band with angular momentum projection $|m|$. The physical interpretation is quite intuitive since the optimal trapping rate is determined by the velocity with which the excitation is reaching the acceptor site, and trapping it too fast leads to the quantum Zeno effect.
\begin{figure*}[ht]
\includegraphics[width=0.8\textwidth]{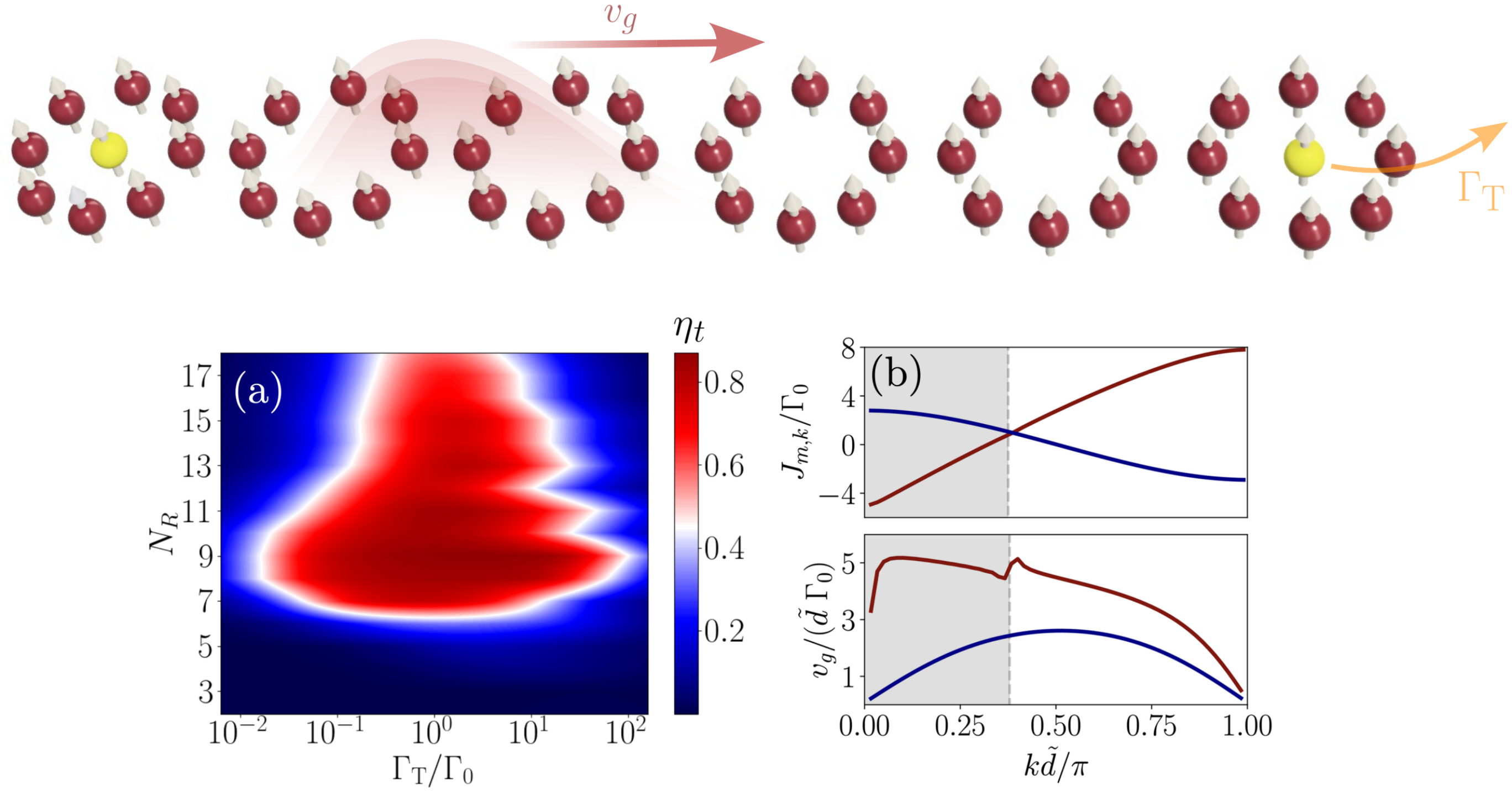}
\caption { \textit{Excitation transport emerges in a limited range of trapping rates.} (a) A numerical scan for the transport efficiency $\eta_t$ over $N_R$ and the trapping rate for 10 rings in a chain. The donor/acceptor detuning $\Delta$ is optimized for each $N_R$. (b) For ordered lattices with subwavelength spacing, the optimal trapping rate is dictated by the group velocity $v_g$ of the collective eigenmode, resonant with the donor/acceptor frequency. Additional parameters: $d_R=d$ for $N_R$ even, $d_R=\sqrt{3}/2 \ d$ for $N_R$ odd and in (a) and (b) a fixed ring radius $R/\lambda_0=0.08$.}
\label{fig5}
\end{figure*}
First, one can already observe the quantum Zeno effect for a single quantum emitter, which features a ground state, an excited state, and an additional trap state $|t\rangle$ to which the excited state decays irreversibly with rate $\Gamma_\mathrm{T}$. Now we add a coherent drive to the system via $\Omega_0 (\hat{\sigma}^\dagger +\hat{\sigma})$. It follows that the effective rate at which excitation is accumulating in the trap state is given by $4 \Omega_0^2 \Gamma_\mathrm{T} /(\Gamma_0 + \Gamma_\mathrm{T} )^2$ in steady state. This effective trapping rate features an optimum at a trapping rate $\Gamma_\mathrm{T}^\mathrm{opt} = \Gamma_0$. This is the well-known quantum Zeno effect for a single driven quantum emitter.
In Fig. \ref{fig5}(a) a numerical scan of the transport efficiently $\eta_t$ after a time $t \Gamma_0 =150$ is shown as a function of the ring emitter number $N_R$ and the trapping rate for 10 rings. Noteably $N_R=8,9,10$ allows for the smallest trapping rate $\Gamma_\mathrm{T}$ where transport is still substantial. In Fig. \ref{fig5}(b) the energy bands ($|m|=2)$ around $J_{m,k}=0$ are shown and the corresponding group velocities $v_g$ plotted. The optimal trap rate is approximately given by the group velocity with which the donor/acceptor emitters are resonant,
\begin{align} \label{trapping}
\Gamma_\mathrm{T}^\mathrm{opt} \approx v_g/\tilde{d},
\end{align}
where $\tilde{d}=2R+d_R$ is the separation between neighbouring ring centers. The group velocity is determined numerically at $J_k = 0$, corresponding to the detuning $\Delta$ of the donor/acceptor emitters. An interesting consequence of Eq. \ref{trapping} is that by controlling the number of emitters $N_R$ per ring, therefore by extension $\tilde{d}$ and the donor/acceptor detuning $\Delta$, a lower optimal trapping rate can be engineered while keeping a high transport efficiency. 

\section*{The single excitation subspace} 

The analysis in the main text is performed in the single-excitation sector for the electronic degrees of freedom. This allows us to rewrite the effective Hamiltonian in non-Hermitian
form as in the main text. The time dynamics follow the Schrödinger equation, $i \partial_t | \Psi(t) \rangle = \hat{\mathcal{H}}_\mathrm{eff}| \Psi(t) \rangle $ with $\hbar = 1$, and the expectation value of observable $\hat{\mathcal{O}}$ becomes ${\mathcal{O}} = \langle \Psi(t) | \hat{\mathcal{O}} | \Psi(t)\rangle$. The advantage is that instead of describing the state of the quantum system by
a density matrix of size $N \times N$ for $N$ emitter,
the system is described in terms of state vectors of size $N$. This makes the numerical analysis feasible for larges system sizes.
The general state vector can be written as
\begin{align}
    |\Psi (t) \rangle = \sum_{m=1}^N \alpha_m(t)  |g,g,...e_m,...\rangle = \sum_{m=1}^N \alpha_m(t)  \hat{\sigma}_m^\dagger  |G\rangle,
\end{align}
 with time-dependent coefficients $\alpha_m(t)$ and $|G\rangle \equiv |g,g,...,g\rangle$ denotes all emitters in the ground state.

\section*{Superradiant corner states in two-dimensional ring lattices}

\begin{figure*}[ht]
\includegraphics[width=0.85\textwidth]{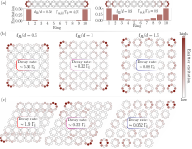}
\caption { \textit{Emerging single-excitation edge and corner states \cite{ota2020active} in one- and two-dimensional quantum emitter ring lattices} (a) Edge states for a 1D chain of rings ($N_R=8$) of various spacings. At the critical spacing $d_R/d\approx 0.58$, the edge states become strongly localized at the edges with vanishing amplitude in the bulk rings. For (b) square lattices and (c) hexagonal lattices with $N_R=8$, corner states become more pronounced and exhibit a transition to superradiant decay. Future studies might show, that these states could exhibit directional coherent light emission. Superradiant corner state lasing in two-dimensional lattices could have the additional advantage of a larger spatial range in absorbing pump light as opposed to edge state lasing in one-dimensional ring chains \cite{st2017lasing}. Emitter spacing is $d/\lambda_0 = 0.05$ in all plots.}
\label{edgestates}
\end{figure*}

The emergence of edge states in a one-dimensional chain of quantum emitter rings is analyzed in the main text. Furthermore, these edge states become superradiant with emission occurring at the ends of the chain. This might hint at the possibility of edge state lasing, observed in chains of golden nano-ring resonators \cite{parto2018edge,st2017lasing}.
Here we show that two-dimensional lattices of rings feature intriguing states, such as 2D edge states and corner states \cite{el2019corner}, illustrated in Fig. \ref{edgestates}. For $N_R=8$ with decreasing inter-ring spacings $d_R$, corner states become more pronounced, exhibiting a transition to superradiance for $d_R \lesssim d$.
The single-excitation corner and edge states are obtained numerically with the quantum optical model described in the main text taking long-range dipole-dipole interactions ($\sim d^{-1}$) into account. By contrast, the observed photonic corner states in kagome lattices \cite{ota2020active} are described via a tight-binding model with exponentially decreasing tunneling strengths $\sim \mathrm{exp}(-d)$ between nearest-neighbor sites. The open quantum system framework described here allows the studying of radiative properties and photon statistics, namely coherent light emission of corner states. We find that edge states with emitter excitation along the whole edge of the 2D lattice are always subradiant, suggesting superradiance in the low-excitation regime as a unique feature of corner states.


\end{document}